\documentclass{PoS}

\title{Search for Low-mass Dark-Sector Gauge Boson with the BABAR Detector}

\ShortTitle{Short Title for header}

\author{\speaker{Romulus Godang}\thanks{BABAR Collaboration}\\
        University of South Alabama\\
        411 University Boulevard, North\\
        Mobile, AL 36688\\
        E-mail: \email{godang@southalabama.edu}}

\abstract{
We report searches for a new muonic dark force mediated by a gauge boson ($Z'$) coupling 
only to the second and third lepton families. The existence of the $Z'$ boson is probed in 
$e^+e^- \to \mu^+\mu^- Z'$, $Z' \to \mu^+ \mu^-$ events, with an analysis based on the full 
data sample collected with the BABAR detector at the PEP-II $e^+e^-$ collider. No significant 
signal is observed. Limits on dark-sector coupling constants are derived; these improve upon 
current bounds, and further constrain the allowed parameter space.
}

\FullConference{38th International Conference on High Energy Physics\\
		3-10 August 2016\\
		Chicago, USA}

\begin{document}

\section{Introduction}
The microscopic nature of dark matter is currently unknown. Models of physics beyond Standard Model 
predict the existence of a new non-Abelian gauge group Higgs with gauge boson masses below 10 GeV~\cite{birkedal}.  
The WIMP hypothesis suggested that dark matter is assumed to consist of stable particle with low masses. Such new gauge 
bosons are typically interact with other Standard Model elementary particles. 
Based on the $L_\mu - L_\tau$ model~\cite{altmannshofer} one of the most promising candidates based on the gauging the existing approximate global 
symmetries of the Standard Model (SM) is the gauge group associated with the difference between muon and tau-lepton number. 
The gauge $L_\mu - L_\tau$ model portal to the $Z'$ has all features of being couples only to the leptons of the second and third generation.  

\section{Data and Event Selection}
We used the data collected by the BABAR detector with the total luminosity of 514 fb$^{-1}$. Most of the data 
were taken at the $\Upsilon(4S)$ resonance including about 28 fb$^{-1}$ data at $\Upsilon(3S)$ and 14 fb$^{-1}$ data 
at $\Upsilon(2S)$ and 48 fb$^{-1}$ data at the off-resonance. The $\Upsilon(4S)$ resonance decays to a pair of 
$\bar{B}B$~\cite{babar05}. We used about 5\% of the data set to validate and 
optimize the analysis method. The data are only examined after finish finalizing the analysis method. For the background study we 
generate signal Monte Carlo (MC) samples.  

Signal MC events are generated using the MadGraph 5~\cite{madgraph5}, which calculates matrix elements for the sample. The MC then 
were showered using the Pythia 6~\cite{pythia6} for a bout 30 different $Z'$ mass hypotheses. The main background comes from the
QED processes. We generate the direct processes of $e^+ e^- \to \mu^+ \mu^- \mu^+ \mu^-$ using the Diag36~\cite{diag36}, which includes
the full set of the lowest order diagrams. Note that Diag36 does not include initial state radiation (ISR) samples. The events 
of the process of $e^+ e^- \to e^+ e^- (\gamma)$ is generated using the BHWIDE~\cite{bhwide} and the MC events of 
$e^+ e^- \to \mu^+ \mu^- (\gamma)$ and $e^+ e^- \to \tau^+ \tau^- (\gamma)$ are generated using the KK~\cite{kk}. The off-resonance 
data samples, $e^+e^- \to \bar{q}q$ (q = u, d, s, c), are simulated using the EvtGen~\cite{evtgen}. The detector acceptance and reconstruction 
efficiency are determined using MC simulation based on GEANT4~\cite{geant4}

\section{Measurement of $Z' \to \mu^+ \mu^-$  }
We select events containing exactly two pairs of oppositely charged tracks, consistent with the topology of the process: 
$e^+ e^- \to \mu^+ \mu^- Z'$ and $Z' \to \mu^+ \mu^-$ final state. The muons are identified by particle identification
algorithms for each track. We require the sum of energies of the electromagnetic clusters that are not associated 
to any track must be less than 200 MeV. We finally reject events that comes from the $\Upsilon(3S)$ and $\Upsilon (2S)$,
where $\Upsilon(2S, 3S) \to \pi^+\pi^-\Upsilon(1S)$, $\Upsilon(1S) \to \mu^+ \mu^-$ decays if the dimuon combination is within
100 MeV of the $\Upsilon(1S)$ where pions are misidentified as muons.

\begin{figure}
\centering
\includegraphics[width=8cm,clip]{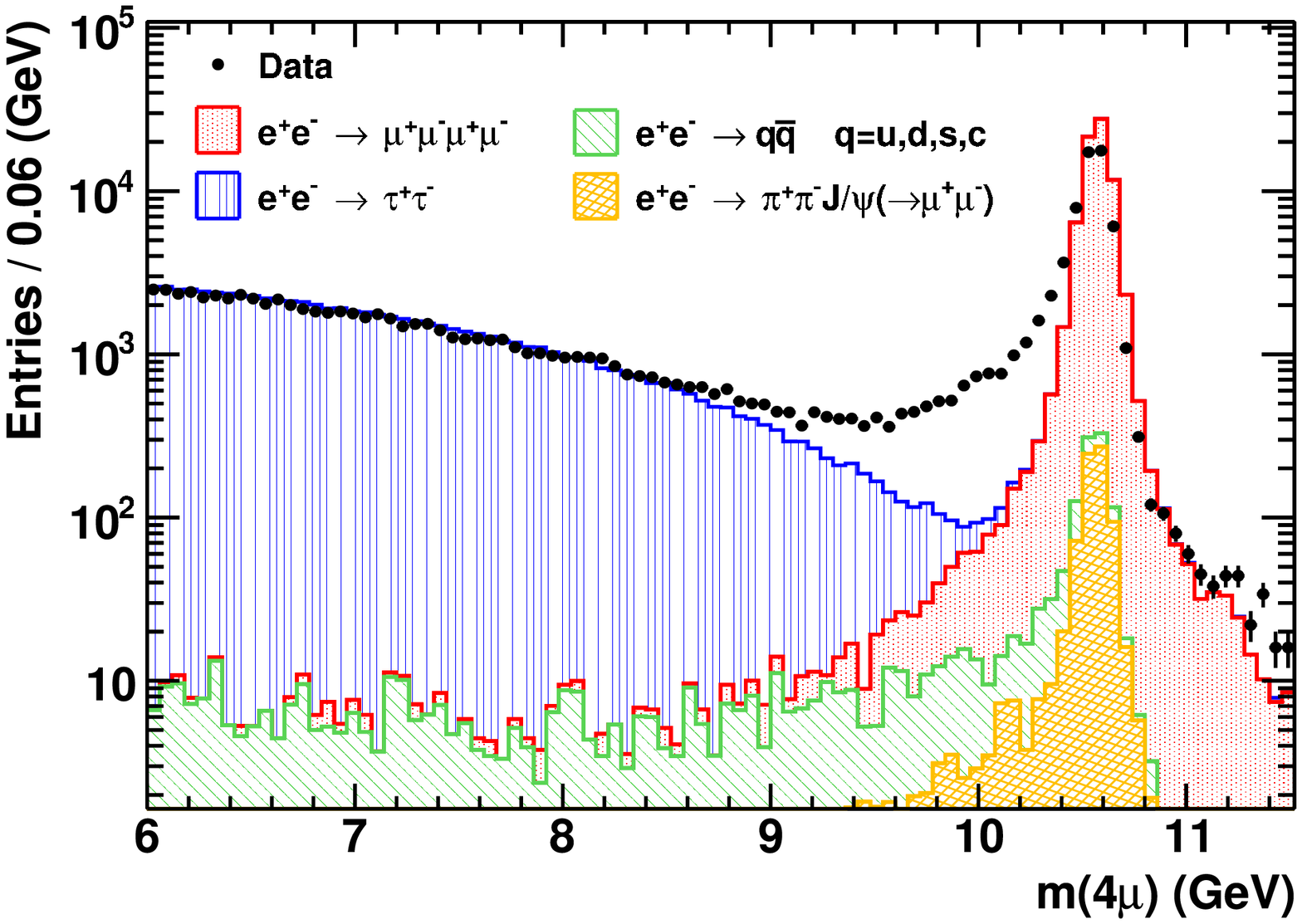}
\caption{The distribution of the reduced dimuon mass, $m_R = \sqrt{m^2_{\mu^+ \mu^-} - 4m^2_\mu}$ in log scale together
with the Monte Carlo predictions of various processes with normalized to the data luminosity.}
\label{fig-2b}     
\end{figure}

\begin{center}
\begin{figure}
\begin{tabular}{lr}
\includegraphics[width=7.5cm]{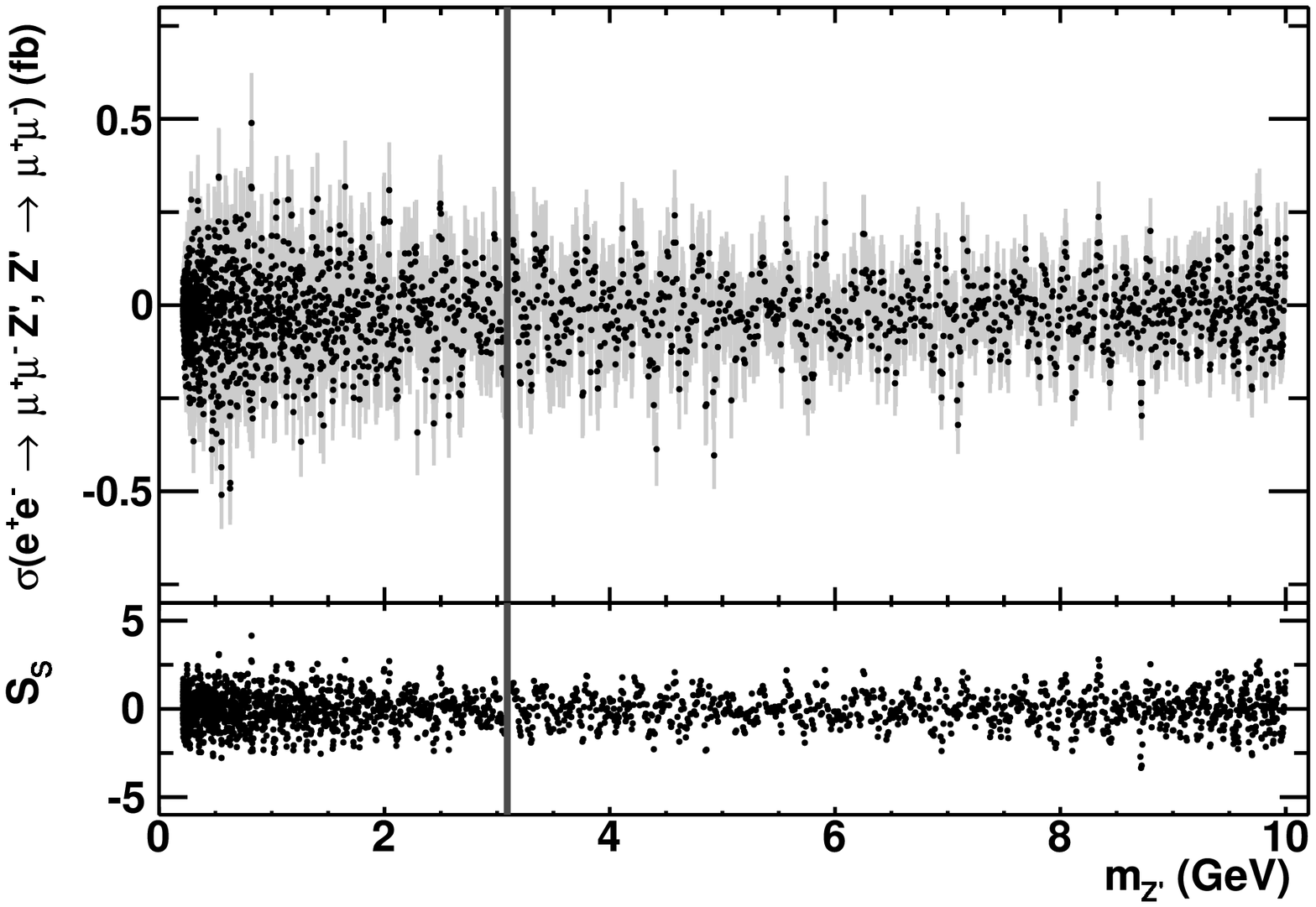}\hspace{0.12cm} 
\includegraphics[width=7.6cm]{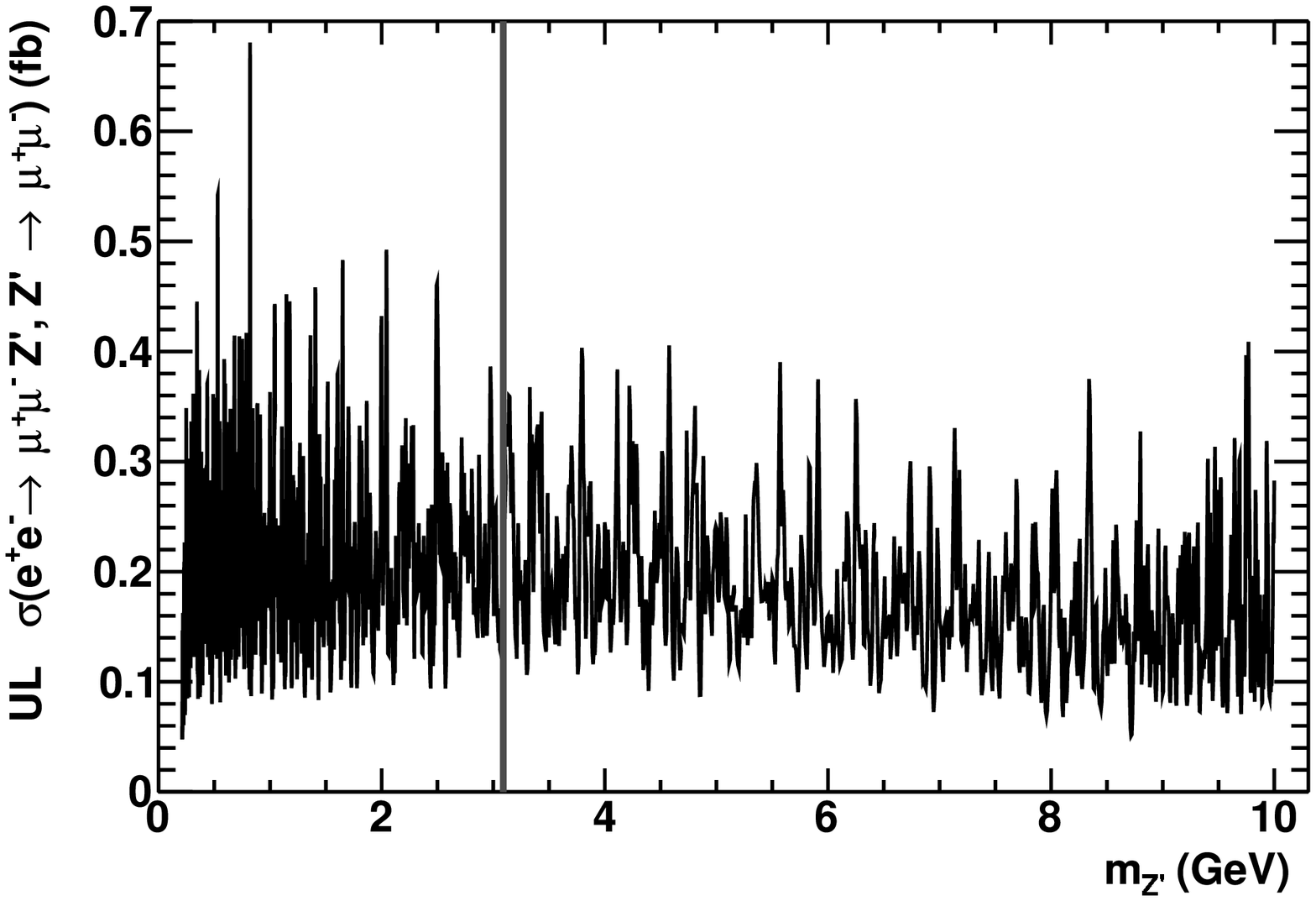}
\end{tabular}
\caption{(Left) The measurement of $e^+ e^- \to \mu^+ \mu^- Z'$, $Z' \to \mu^+ \mu^-$ cross section with its statistical
significance as a function of the $Z'$ mass. The excluded region is indicated by the gray band.
(Right) The limit on the cross section $\sigma (e^+ e^- \to \mu^+ \mu^- Z'$, $Z' \to \mu^+ \mu^-)$  as a function 
of the $Z'$ mass. The excluded region is indicated by the gray band. }
\label{fig34}
\end{figure}
\end{center}

The reduced dimuon mass is calculated using the following equation
$m_R = \sqrt{m^2_{\mu^+ \mu^-} - 4m^2_\mu}$ in log scale is shown in Fig.~\ref{fig-2b}. The most dominant samples is coming from the direct 
decay of $e^+ e^- \to \mu^+ \mu^- \mu^+ \mu^-$ process. The contribution from the decay of $\Upsilon(2S) \to \pi^+ \pi^- J/\psi$, $J/\psi \to 
\mu^+ \mu^-$ as shown around 3 GeV. 
The signal efficiency at low masses is about 35\% and it rises to about 50\% around higher mass of the reduced dimuon mass. 
We exclude the $J/\psi$ region when calculating the correction factors by fitting the simulated and reconstructed reduced 
dimuon masses in the range of $1 < m_R < 9$ GeV.
The signal yield is extracted bu a series of unbinned likelihood fits to the reduced dimuon mass spectrum within the range of $0.212 < m_R < 10$ GeV 
and $0.212 < m_R < 9$ GeV for the $\Upsilon(4S)$ resonance data and $\Upsilon(2S)$ and $\Upsilon(3S)$ resonances data, respectively.  
We exclude a region of $\pm 30$ MeV around the nominal known $J/\psi$ mass. We probe a total of 2219 mass hypothesis. The cross section of
$e^+ e^- \to \mu^+ \mu^- Z'$, $Z' \to \mu^+ \mu^-$ is extracted as a function of $Z'$ mass as shown in Fig.~\ref{fig34} (left). The gray band indicates 
the excluded region. We find the largest local significance is $4.3 \sigma$ around $Z'$ mass of 8.2 GeV that is corresponding to the global
significance of $1.6 \sigma$ and it is consistent with the zero-hypothesis.  
We also derive $90\%$ confidence level (CL) Bayesian upper limit on the cross section of $e^+ e^- \to \mu^+ \mu^- Z'$, $Z' \to \mu^+ \mu^-$ as shown in
Fig.~\ref{fig34} (right). 

\begin{figure}
\centering
\includegraphics[width=10cm,clip]{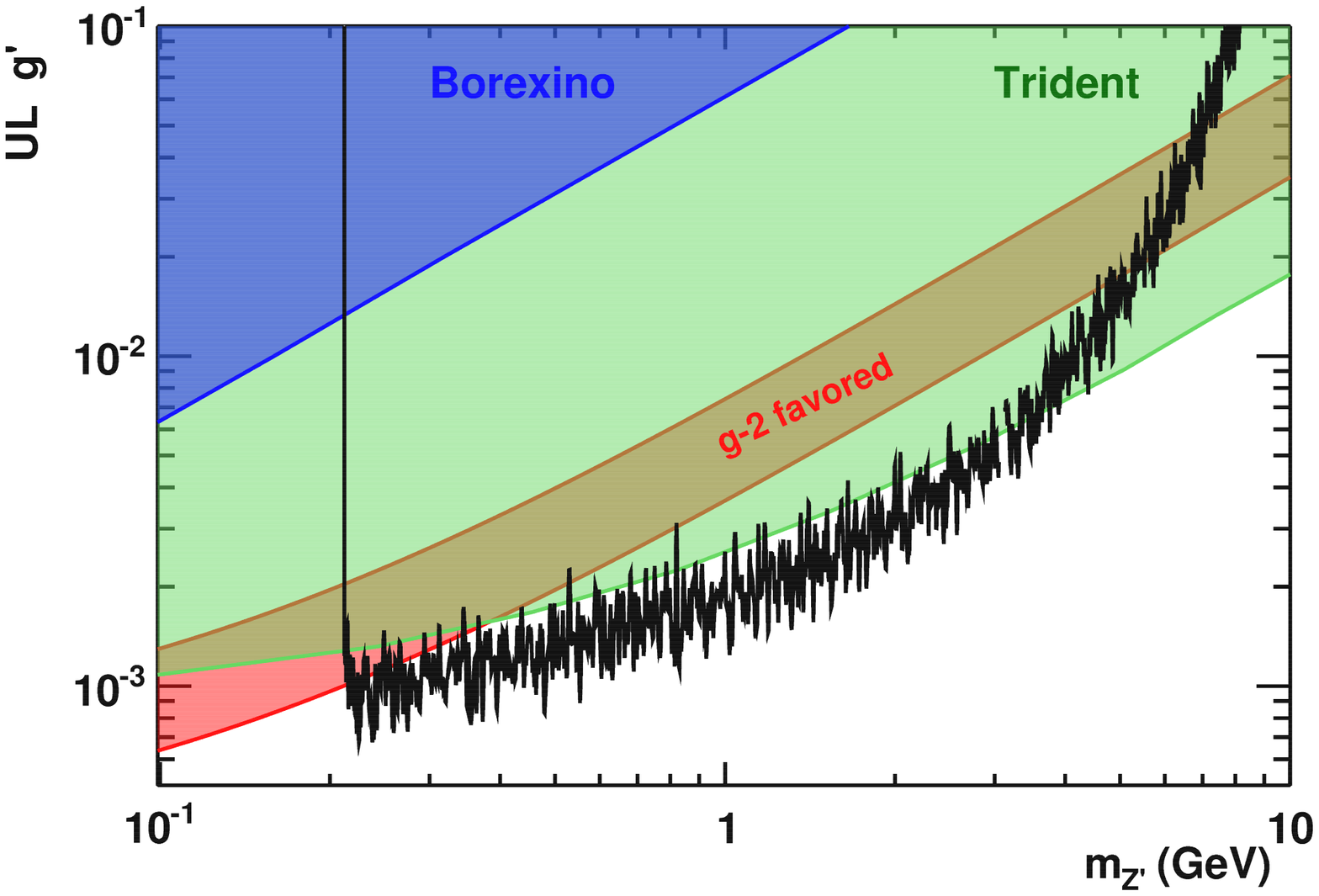}
\caption{Upper limit on the new gauge coupling $g'$ as a function of the mass of $Z'$ together with
the existing limits from Borexino and neutrino experiments.}
\label{fig-05}      
\end{figure}

We consider all uncertainties to be uncorrelated except for the uncertainties of the luminosity and efficiency. We finally extract
the corresponding $90 \%$ CL on the coupling parameter $g'$ by assuming the equal magnitude vector couplings muons, taus and the corresponding
neutrinos together with the existing limits from Borexino and neutrino experiments as shown in Fig.~\ref{fig-05}.  
We set down to $7 \times 10^{-4}$ near the dimuon threshold.

\section{Conclusion}
In conclusion, we have performed the first direct measurement of $Z'$ production from the decay of $e^+ e^- \to \mu^+ \mu^- Z'$, 
$Z' \to \mu^+ \mu^-$ an $e^+ e^-$ collider at BABAR.  No significant signal is observed for $Z'$ masses in the range of
0.212 - 10 GeV. We set limits on the coupling parameters $g'$ down to $7 \times 10^{-4}$. We set a strongest bounds for many parameter 
space below 3 GeV. We exclude most of the remaining [parameter space preferred by the discrepancy between the calculated and measured 
anomalous magnetic moment of the muon above the dimuon threshold~\cite{echenard}.

The author would like to thank the organizers of the $38^{th}$ International Conference on High Energy Physics, Chicago, United States of America,
the BABAR Collaboration, the University of South Alabama, and the University of Mississippi.

\end{document}